\documentstyle[aps,pre,floats,epsf]{revtex}

\begin{document}
%\draft
\twocolumn[\hsize\textwidth\columnwidth\hsize\csname@twocolumnfalse%
\endcsname 

\title{Multicanonical methods vs. Molecular Dynamics
vs. Monte Carlo: Comparison for Lennard--Jones glasses.}

\author{Kamal K. Bhattacharya\thanks{kamal@msc.cornell.edu} and James P. Sethna\thanks{sethna@msc.cornell.edu}} \address{Laboratory
of Atomic and Solid State Physics, Cornell University, Ithaca, NY
14853-2501}

\date{\today}
\maketitle

\begin{abstract} 
We applied a multicanonical algorithm (entropic sampling) to a
two--dimensional and a three-dimensional Lennard--Jones system with
quasicrystalline and glassy ground states. Focusing on the ability of
the algorithm to locate low lying energy states, we compared the
results of the multicanonical simulations with standard Monte Carlo
simulated annealing and molecular dynamics methods. We find slight
benefits to using entropic sampling in small systems (less than 80
particles), which disappear with larger systems. This is disappointing
as the multicanonical methods are designed to surmount energy barriers
to relaxation. We analyze this failure theoretically, and show (1) the
multicanonical method is reduced in the thermodynamic limit (large
systems) to an effective Monte Carlo simulated annealing with a random
temperature vs. time, and (2) the multicanonical method gets trapped
by unphysical entropy barriers in the same metastable states whose
energy barriers trap the traditional quenches. The performance of
Monte Carlo and molecular dynamics quenches were remarkably similar.
\end{abstract}
]

\section{Introduction}

In the past decade there has been an outpouring of interest in
accelerating statistical mechanics simulations.  This started with the
work of Swendsen and collaborators: Swendsen and Wang introduced a
cluster--flip method for accelerating non-disordered spin systems
\cite{sw86,sw87}, and Widom, Strandburg, and Swendsen introduced a
cluster--flip for finding quasicrystalline ground--states in a
two--dimensional atomic simulation \cite{wss87}. These methods all
gain a major speedup by introducing mostly non-local update rules, and
often prove capable of bypassing critical slowing down problems
\cite{swf92}.

During the same period, different accelerating approaches were
introduced, which are based on efficient schemes to analyze data from
traditional Monte Carlo simulations \cite{swf92,fs88,lk90} and are
frequently called ``histogram methods''. These methods have enlarged
the applicability of various kinds of critical phenomena simulations,
although they are not necessarily designed to bypass critical slowing
down problems as efficiently as {\it e.g.} cluster
algorithms. Nevertheless, substantial progress can be achieved
combining histogram and cluster--flip algorithms (see {\it e.g.}
reference \cite{nb96}).

More recently so--called ``reweighting techniques'' have been
introduced, which are based on an early approach by G. M. Torrie and
J. P. Valleau \cite{tv77}. They proposed a method to enlarge the
sampling range of a Monte Carlo algorithm by using nonphysical
weighting functions. The general idea in the newer approaches is to
change the relative weights of different configurations to sample
equally in all ranges of energy rather than focusing on a narrow
temperature range. The most frequently used reweighting method is
multicanonical sampling \cite{bn91,bn92,b97,j94,sb95}, which
represents the most general method as other reweighting methods, {\it
e.g.}  entropic sampling \cite{l93}, can be directly mapped onto this
approach\cite{bho95}. In systems with a strongly double peaked
probability distribution of magnetization or energy states (a
situation often found in systems exhibiting a first--order phase
transition), the multicanonical approach has been proven to be a
powerful tool.  Simple reweighting schemes allow to overcome the
``supercritical slowing down''\cite{j94} known from canonical Monte
Carlo simulations at a fixed temperature. 
{\it E.g.} in non-disordered spin systems with a field-driven
first-order phase transition ({\it e.g.} the Ising model) or a
temperature-driven first-order phase transition ({\it e.g} the q-state
Potts model) the supercritical slowing down of canonical Monte Carlo
is due to the low Boltzmann weight of the domain-wall
states. Multicanonical sampling approaches the problem with
introducing a weightfunction, which weights all magnetization states
(Ising model) or energy states (Potts model) equally, and therefore
ensures that domain-wall states are sampled with the same likelihood
as all other accessible states. The canonical distribution function at
a fixed temperature, which contains all the thermodynamic information,
can be reconstructed. Usually multicanonical sampling uses local
update schemes along the lines of the Metropolis
algorithm\cite{mrrtt53}; variations using cluster--flip or other
methods are feasible and have been proven useful (for a review consult
\cite{b97,j94} and references therein).

Of course, acceleration methods are most crucial for glassy systems,
which otherwise can be inaccessible to numerical
simulations\footnote{Even the experiments fall out of equilibrium.
Think of an experiment as $10^{23}$ parallel atomistic processors with
picosecond clock times!}.  Whether one believes that glasses are
sluggish because of large energy barriers to relaxation (rates $\sim
e^{B/T}$), or believe that the free energy barriers are due to
tortuous entropically difficult routes between the metastable
configurations, a clever algorithm could in principle jump directly
between the glassy states.  Instead of relative rates which grow as a
power of $T-T_c$, acceleration could gain us exponential speedups.

Acceleration methods have been extensively applied to disordered spin
models.  These studies have been less focused on understanding the
performance of the algorithms, because the physics of the systems is
less thoroughly understood (there has been more to mine from the
results, and a less firm foundation on which to do algorithmic
analysis). The multicanonical methods have been applied to spin
glasses in two and three dimensions to calculate the zero--temperature
entropy, ground state energies, distribution of overlaps etc. (see
refs. \cite{spin1,spin2,spin3,spin4,spin5,spin6,spin7,spin8,spin9}). The
authors succeed in evaluating these properties with remarkable
accuracy; nevertheless, whether the replica theory \cite{mpv87} or the
droplet scaling ansatz \cite{fh88} is the more appropriate picture in
describing the ground state properties of glasses could not be
resolved. The performance of multicanonical sampling for glassy
systems is clearly worse than for system with a less rugged
landscape. We believe this failure is systematic and can't be avoided
within the framework of multicanonical sampling.

The authors of reference \cite{spin1} argued multicanonical methods
should be superior to simulated annealing (gradual
cooling)\cite{kgv83}, although a direct comparison was made only to
canonical sampling (quenches to a fixed low temperature). The authors
of ref. \cite{lc94} applied multicanonical sampling to the Traveling
Salesman problem and claim to acchieve a dramatic improvement over the
traditional Monte Carlo simulated annealing approach. Newman\cite{n97}
has used both cluster methods\cite{nb96} and entropic sampling
\cite{l93} to study the random--field Ising model. He finds dramatic
speedups from both methods, often reaching equilibrium in a few passes
through the lattice.  Newman has focused on small systems (mostly
$24^3$, with a few runs for systems up to $64^3$), and simultaneously
used histogram methods to measures critical exponents and phase
boundaries for a range of disorders and temperatures. He confirms
results of the related ``simulated tempering'' approach, invented by
E. Marinari and G. Parisi \cite{mp92,m96}. Simulated tempering proved
very useful in spin glass simulations \cite{kr94} and is similar in
spirit to the multicanonical approach \cite{ho96}.

Acceleration methods have been little used in continuum atomic
simulations, perhaps because of the widespread reliance on molecular
dynamics methods. Straightforward, direct molecular dynamics
simulations of the equations of motion do not converge to an
equilibrium state much faster than the Monte Carlo simulated annealing
methods, but they are also not noticeably worse\cite{c97}, and they
have a direct physical interpretation.  Shumway\cite{ss91} studied a
one-dimensional atomic system in an incommensurate sinusoidal
potential, and developed an evolutionary algorithm which generated
optimal cluster moves as the system was quenched to lower
temperatures; later attempts to generalize these ideas to higher
dimensions have so far not been successful\cite{s97}. The authors of
reference \cite{ho94} used multicanonical sampling and Monte Carlo
simulated annealing to study the folding of the peptide
Met-enkephalin; the multicanonical method found the ground state more
consistently using the same amount of computer time. This result
underlines the general belief \cite{spin1,b96} that simulations in the
multicanonical ensemble are in many ways superior to traditional
simulated annealing.

In this paper we apply multicanonical sampling in the particular form
of entropic sampling to two-component Lennard--Jones systems, and
compare the performance with traditional simulated annealing and
straightforward molecular dynamics in finding low energy
configurations. We search for low energy states of a three dimensional
Lennard--Jones glass, one of the prototype glassy
systems\cite{sw84,sw89,sw85,ws85}, and use the set of parameters
recently introduced by W. Kob and H. C. Andersen
\cite{ka94,ka951,ka952}. In addition to that, we apply entropic
sampling and simulated annealing to a two-dimensional Lennard--Jones
systems with quasicrystalline ground states, using the parameters of
reference \cite{wss87}.  We find that entropic sampling brings little
benefit for the study of either. We argue that this is likely a
general effect, applicable to all simulation methods applied to glassy
systems in the thermodynamic limit.

\section{Introduction to the Methods: Multicanonical Sampling and 
Entropic Sampling}

The standard way of implementing a Monte Carlo algorithm is
using importance sampling. The idea behind this approach is
simple. Rather than weighting each sample in phase space equally, each
state is weighted with a sample probability distribution
$\Gamma(x)$, where $x$ denotes the sampled configuration of the
system.  To estimate the thermal
average of an observable $A$, one calculates:

\begin{equation}
<A> = \frac{\sum_x A(x) \exp[-\beta H(x)] \Gamma^{-1}(x)}{\sum_x
	\exp[-\beta H(x)] \Gamma^{-1}(x)} \; ,
\end{equation}
where $H$ is the Hamiltonian of the system (so $H(x)$ is the energy
$E$ for the state $x$) and $\beta = 1/k_B T$.  Choosing $\Gamma(x)$
non-uniformly ensures that states with important contributions to the
partition sum are preferentially sampled, and therefore the number of
states need to be sampled to provide a reasonable estimate of $A$ is
significantly reduced.

In standard Monte Carlo methods, {\it i.e.} canonical Monte Carlo or
simulated annealing, the weighting distribution is the Boltzmann
distribution $\Gamma =\exp[-\beta H(x)]$.  This has the advantage of a
direct physical interpretation: the computer is doing the same thermal
average as an equilibrium system at temperature $1/(k_B \beta)$.  It
has an important disadvantage that configurations and events which are
rare in the physical system are also rare in the simulation.  In
particular, if the system has a ``rugged energy landscape'', with
large free energy barriers $B$ separating physically important
metastable states, the system will cross between these states with the
same slow Arrhenius rate $\nu \exp{(-B/T)}$ that is found
experimentally.

The idea of multicanonical sampling is to circumvent this problem by
choosing $\Gamma(x)$ so that the distribution of states
$P(H(x))~\sim~\Omega(H(x))~\times~\Gamma(x)$ is approximately flat in
energy (or some other variable, like magnetization\cite{j94}).  In
principle, we want to choose $\Gamma(E) = 1/\Omega(E) = \exp[-S(E)]$,
where $\Omega(E)$ is the density of states at energy $E$ and $S(E)$ is
the entropy.  Of course, we don't begin the simulation knowing the
entropy as a function of energy!

In our work we use the entropic sampling algorithm \cite{l93}, which
is a numerical and mathematical equivalent variant of the
multicanonical approach \cite{bho95}. The only difference between
entropic and multicanonical sampling is the way by which one generates
estimates $J(E)$ of $S(E)$.  The entropic sampling algorithm uses a
quite straightforward recursive updating method:

\begin{itemize}
\item[1:] Initialize to zero an array $H(E)$, which will keep track of the
energy of the visited states.
\item[2:] Sample states according to the current $J_i(E)$, and add the
energy of each sampled state to the histogram $H$, for a reasonably
long time.
\item[3:] Set the new $J_{i+1}$ according to the following rule:
\noindent
\begin{equation}
      J_{i+1}(E) = \left\{\begin{array}{ll}
                        J_i(E) + \log(H(E)), & {\rm if} H(E) \neq 0 \\
			J_i(E), &  {\rm if} H(E) = 0 
                      \end{array}\right.
\end{equation}
\end{itemize}

\noindent

The multicanonical sampling update scheme differs from entropic
sampling in the treatment of the histogram bins with few entries (for
an analysis of various schemes see references \cite{sb95} and
\cite{b96}).  The original approach introduces a constant  slope
for $J(E)$ below a cutoff energy, corresponding to a small constant
temperature.  These extra parameters are annoying \cite{b96} in the
implementation; however, they do tend to keep the system from being
trapped in energy regions which have not hitherto been sampled
frequently.  As we will argue, in glassy systems both algorithms will
tend to get trapped in low energy metastable states even when the
statistics are fine\footnote{In any case our simulations spend around
half the time at high energies, so any algorithmic improvements can
bring at best a factor of two in computer time.}.  In this paper we
use the simpler entropic sampling method of equation~(2).

\section{Theoretical Expectations for Relative Performance}

What makes people think that multicanonical sampling should be an
improvement over simulated annealing or molecular dynamics?  We consider
three possible reasons. 

(1)~Perhaps the multicanonical method is better because it allows the
system to cross energy barriers (as is mentioned frequently
\cite{bn91}--\cite{l93})?  This is indeed an improvement over
canonical sampling at a fixed temperature; however, a simulated
annealing method also runs at a variety of temperatures.

Indeed, the two methods are {\it identical} in the thermodynamic limit.
The acceptance ratio for a given single--atom Monte Carlo move for 
entropic sampling is $P(E) =\exp[S(E) - S(E')]$.  In a large system with 
$N$ atoms, the entropy density $S(E)/N$ is a smooth function of the
energy density $E/N$; since the energy density change for a single-atom
move $(E'-E)/N$ is small, we may expand $S(E)$ to first order in $E'-E$.
Using the relation $\partial S(E)/\partial E = 1/T$, the acceptance ratio 
becomes $P(E) = \exp[-(E'-E)/T]$.  Thus entropic sampling at the energy 
$E$ has exactly the same acceptance ratio as simulated annealing at a 
temperature $T(E) = (\partial S(E)/\partial E)^{-1}$.

Thus the local behavior --- the acceptance ratio for Monte Carlo moves
from a given state --- is virtually the same for multicanonical and
canonical sampling\footnote{Near first-order transitions, canonical
quenches produce large changes in the state for small changes in
temperature, and thus behave quite differently from the multicanonical
approaches (which by varying the energy explore the interface states
directly).  This is one of the major applications of multicanonical
sampling methods.  We expect that the multicanonical methods will
perform for these systems rather similarly to microcanonical quenches
which conserve the energy; see ref. \cite{nsskn94}. In our glassy
simulations, this distinction is presumably not important.}. The
differences between the two methods near a given state should be
similar in magnitude and type to the differences between the
microcanonical (fixed-energy) simulations and the canonical
(fixed-temperature) simulations: differences can be seen for small
systems, but disappear as the system gets larger.  To be explicit, for
a large system the final state of an entropic sampling run for which
the time-dependent energy is $E(t)$ should be statistically equivalent
to a simulated annealing run with randomly fluctuating temperature
$T(E(t))$. One notes also that the quench rate is not tunable for the
multicanonical method: the ``diffusion constant'' in energy space
depends on the atomic step--size and on the number of particles. This
is potentially a serious handicap, as changing the quench rate is the
primary tool used in glassy systems to find lower energy states.

Thus the power of multicanonical methods to vary the energy to
facilitate barrier crossing is --- for large systems at least --- no
different from repeatedly heating and cooling the entire system.

(2)~Perhaps the multicanonical method might be picking the heating and
cooling schedule intelligently, in order to escape from local minima?
Indeed, since the effective temperature becomes lower as the energy
decreases, an entropic sampling system stuck in a high--energy
metastable state will have larger thermal excitations (bigger
acceptance of upward moves in energy) than one in a low energy state,
and will depart faster.  This is the explanation, we believe, for the
substantial success of the entropic (multicanonical) sampling method
seen in the past.

This preferential escape from high--energy metastable states will
unfortunately also become unimportant for large systems.  One can see
this most easily by considering a local region trapped in a
high-energy configuration with local energy $e'$, and with a lower
energy configuration $e$ nearby, separated by a barrier $b$.  For a
small system, where the local energy difference $e'-e$ is important,
the effective temperatures in states $e'$ and $e$ will differ, but for
a large system of size $N$ this temperature difference (from the
differing acceptance ratios from the two states) will vanish as $1/N$.
There are glassy systems in which the energy barriers and energy
differences are not all local: mean-field spin glasses, for example,
have energy barriers which grow as powers of the number of spins $N$
\cite{mpv87}.  However, the maximum energy barrier (and presumably the
maximum energy asymmetry $e'-e$) scales with a power $N^\alpha$ with
$\alpha$ strictly less than one (at least in finite dimensions), so
the change in effective temperature $N^{\alpha}/N$ still vanishes as
$N\to\infty$\cite{s81}.

(3)~Perhaps the multicanonical method is exploring different energies
more effectively than an externally chosen cooling schedule for
simulated annealing?  For example, the multicanonical method is
guaranteed to converge to an equilibrium density of states at each energy.
The same is true for a simulated annealing run at infinitely slow cooling,
but is not true for repeated coolings at a fixed rate, which would 
be expected to generate metastable states repeatedly.  On the
one hand, the theorems suggest that the ground state should be occupied
as often as any other energy; on the other hand it is hard to see how
a multicanonical quench to low energies can bypass the metastable
states that trap simulated annealing runs of comparable computer time.

To address this issue, let us consider the characteristics of the
random walk $E(t)$ that the system performs in energy space as a
function of time within the multicanonical approach.  For some
systems, such as the Ising model, multicanonical sampling does indeed
produce a roughly unbiased random walk (if one starts from a good
estimate of the density of states).  As the system becomes larger, the
energy range scales with $N$ and the step-size of the energy stays
fixed, so the time--scale for diffusing from high energies to near the
ground--state scales as $N^2$ (distance scales with the square root of
time).  This behavior is confirmed in simulations\cite{bn91,b97}
studying, {\it e.g.}  the first--order phase transition below $T_c$
changing the Ising model from pointing up to down, where traditional
canonical methods would suffer from the surface tension barrier
$\sigma L^{d-1} = \sigma N^{(d-1)/d}$ and so the time scales as $N^2
\exp{\sigma N^{(d-1)/d}}$.  Bypassing this ``supercritical
slowing--down''\cite{j94} is an important application for
multicanonical methods.

It is known numerically that this simple argument breaks down in
simulations of spin glasses\cite{spin1}. The typical time to cover the
energy range (called the ergodicity or tunneling time in the
literature) for spin glasses scales as $N^4$\cite{spin3} or perhaps
$e^N$ \cite{b97} instead of $N^2$.  Why should the random walk
argument not work for glasses?

%FIGURE 1
\begin{figure}[htbp]
\begin{center}
\leavevmode
\epsfysize=7cm \epsfbox{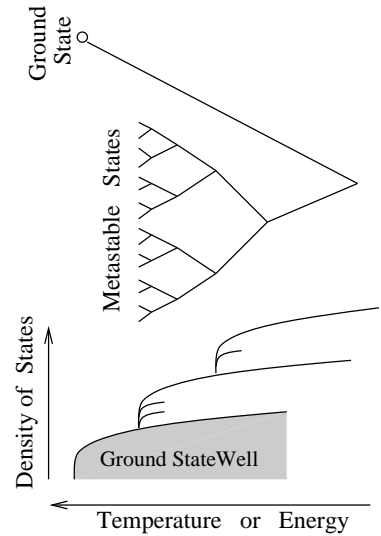}
\vspace{1cm}
\caption{
The glassy metastable states are often thought to form a tree-like
 structure. We anticipate that the inaccessible metastable states will
 form a barrier to leaving the ground state within the multicanonical
 approach, because the total density of states grows much faster than
 the accessible density of states. }
\end{center}
\label{fig1}
\end{figure}

The answer to why the random walk argument breaks down, we assert, can
be found in the trapping of entropic sampling due to the same
metastable states explored in thermal coolings.  Indeed, it is the
metastable states that the system is {\it not able to explore} that
trap the entropic sampling algorithm.  The states of glassy systems
are often described in a caricature tree-like structure (sideways in
figure~1).  The horizontal axis of the tree can be thought of as
either energy or temperature: the branches represent mutually
inaccessible ergodic components.  For the Ising model, there are two
major ergodic components (corresponding to the two directions for the
magnetization) and a few domain-wall states.  For glasses, the ergodic
components are sometimes thought of as regions of configuration space
separated by infinite free energy barriers (as in the mean-field spin
glass models\cite{mpv87}), and sometimes as regions separated by
energy barriers which are too large to cross in the time--scale of the
experiment or simulation.

The key point is that the accessible density of states for a glass can
be very different from the total density of states.  In figure~1, we
note that the ergodic component containing the ground state has a
density of states which differs from the density of states for the
system as a whole, starting at the energy of the first accessible
metastable states.  The number $\Sigma$ of these inaccessible
metastable states is related to the density of tunneling states in
configurational glasses\cite{zp71,ahv72,p81}, and is thought to
increase exponentially with the size of the system ($M$ independent
two--state systems with uncrossable barriers generate $\Sigma = 2^M$
states).  In spin glasses, the number of components separated by
infinite free energy barriers (ones which diverge as $N\to\infty$)
diverges with a power of $N$\cite{s81}.

%FIGURE 2
\begin{figure}[htbp]
\begin{center}
\leavevmode
\epsfysize=7cm \epsfbox{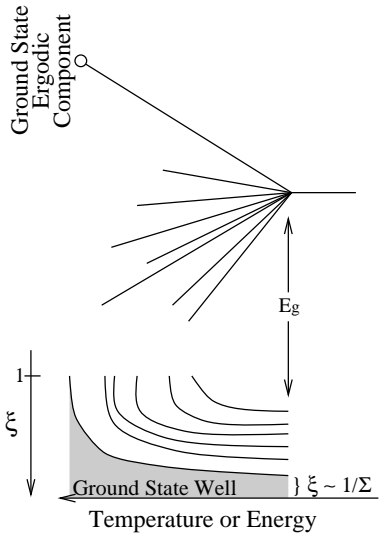}
\vspace{1cm}
\caption{
Grossly oversimplified picture of of glassy states, assuming that all
components merge at the glass transition at energy $E_g$. The
parameter $\xi$ denotes the fraction of ergodic components at each
energy or temperature. At the glass transition, the fraction of
accessible ergodic components is of order $1/\Sigma$, thus the time to
find the ground state ergodic component (or leave it once found)
scales with $\Sigma$ as $\Sigma\to\infty$.}

\end{center}
\label{fig2}
\end{figure}

The multicanonical sampling method is guaranteed to sample the
ground-state energy just as much as any other energy.  It is easy to
see, however, that once the system is in the ground state, it will
stay there for a long time!  If the density of states within the
ground--state ergodic component is $\tilde\Omega_g(E)$ and the density
of states in the entire system is $\Omega(E)$, then the acceptance
ratio for a multicanonical sampling move from $E$ to $E'$ is
$\Omega(E)/\Omega(E')$, while the probability of a random move raising
the energy is $\tilde\Omega_g(E')/\tilde\Omega_g(E)$.  Hence the
likelihood of sampling high energy states $E$ within the ground-state
component will fall as $\tilde\Omega_g(E)/\Omega(E)$.  Consider the
very crude model where all ergodic components are similar and stay
completely separate until the glass transition energy $E_g$ (the
energy at the glass transition temperature), at which point they merge
(figure 2).  For this model, escaping from the ground state component
will take a time which scales as the total number of ergodic
components, and hence diverges as $N\to\infty$. Since multicanonical
sampling spends the same amount of time in each energy range, the time
between independent visits to the true ground state will scale as the
time needed to escape from the ground state ergodic component.

So, we begin our exploration with the expectation that multicanonical
sampling should be useful for small systems, but will not provide
significant advantages for large system sizes.

\section{Implementation}

We argued in the previous section that entropic sampling will not be a
fundamental improvement over repeated coolings using simulated
annealing, at least in large systems.  On the other hand, entropic
sampling and other multicanonical methods have been reported to lead
to substantial gains in equilibration times for small glassy spin
systems\cite{spin1,spin2,spin3,spin4,spin5,spin6,spin7,spin8,spin9}.
We are interested in simulations of structural glasses: collections of
atoms which typically form metastable, glassy configurations when
slowly cooled.  In this section, we give a detailed description of our
implementation of entropic sampling, simulated annealing, and
molecular dynamics.  In order to ensure a fair comparison, we have as
far as possible taken cooling schedules and time and spatial step
sizes from standard references in the literature.

In our three dimensional simulations, we applied the three algorithms to
a binary mixture of large (L) and small (S) particles with the same
mass, interacting via the Lennard--Jones potential of the form
$V_{\alpha\beta}(r)=4\,
\epsilon_{\alpha\beta}[(\sigma_{\alpha\beta}/r)^{12} -
(\sigma_{\alpha\beta}/r)^{6}]$. The values of $\epsilon$ and $\sigma$
were chosen as follows: $\epsilon_{LL} = 1.0, \sigma_{LL} = 1.0,
\epsilon_{LS} = 1.5,\sigma_{LS} = 0.8, \epsilon_{SS} = 0.5, \sigma_{SS}
= 0.88$. All results are given in reduced units, where $\sigma_{LL}$
was used as the length unit and $\epsilon_{LL}$ as the energy
unit. The systems were kept at a fixed density ($\rho\approx 1.2$),
periodic boundary conditions have been applied and the potential has
been truncated appropriately according to the minimum image
rule\cite{ta86}, and shifted to zero at the respective cutoff.  The
minimum image rule prevents a particle from using the periodic
boundary conditions to see more than one copy of its neighbors: to use
the conventional cutoff at $r=2.5\sigma$ would demand at least 160
particles.  The choice of parameters follows recent simulations of
Lennard Jones glasses \cite{ka94,ka951,ka952,vkb96}; this choice
suppresses recrystallization of the system on molecular dynamics time
scales. This potential together with this set of parameters mimics the
potential for ${\rm Ni_{80}P_{20}}$. We looked at 5 different system
sizes ($N$=20, 40, 60, 80, 100). For each $N$ we generated 30 low
energy configurations. The initial configurations were random in the
case of simulated annealing and entropic sampling, and high
temperature equilibrium configurations in the molecular dynamics
case. To compare the three methods, we defined one run length to be
$10^6$ sweeps through the system for the two Monte Carlo methods. The
molecular dynamics runs are quenched at a rate which consumes the same
CPU time as used by the Monte Carlo sampling.

In our two dimensional simulations, we did not apply molecular
dynamics (our hard--wall boundary conditions made it inconvenient).
We again used a binary Lennard Jones system, introduced by Widom,
Strandburg, and Swendsen \cite{wss87} with a slightly unconventional
form for the potential:
$V_{\alpha\beta}(r)=\epsilon_{\alpha\beta}[(\sigma_{\alpha\beta}/r)^{12}
- 2 (\sigma_{\alpha\beta}/r)^{6}]$.  The Lennard Jones parameters are
chosen to favor configurations of decagonal order ($\epsilon_{LL} =
0.5, \sigma_{LL} = 1.176,
\epsilon_{LS} = 1.0,\sigma_{LS} = 1.0, \epsilon_{SS} = 0.5,
\sigma_{SS} = 0.618$), and the system is known to have a quasicrystalline
ground state. The particles are initially randomly distributed in a
large cylindric box with infinitely high walls. The potential was
truncated at $r_{cutoff} = 2.5\sigma_{\alpha\beta}$ and shifted to
zero at this point. All results are in reduced units with
$\epsilon_{LS}$ and $\sigma_{LS}$ as fundamental units.  Here 4
different system sizes (N=31, 66, 101, 160) were used, where the
number of different particles were chosen to keep the ratio fixed
close to the value of 1.06 large atoms per small atom. The authors of
reference \cite{wss87} found that this ratio led to defect--free
ground states.  This system provides an excellent testing ground for
entropic sampling for various reasons. The ground state is known to be
quasicrystalline, a state with a strong bond orientational order
without a long-range order periodicity. Defective configurations are
easily recognized, as the typical defects consists of triangles of
like particles. There are plenty of metastable states with high energy
barriers, as it takes rearrangement of a large number of particles to
disentangle the triangular defects. The authors of reference
\cite{wss87} have shown that simulated annealing fails to locate
ground states, and always gets trapped in a long lived metastables
states, a problem they circumvented using three--particle
cluster--flips.

The final minimum energy configurations for runs of all three methods
were optimized by starting from the lowest energy configurations
found, and quenching down to $T=0$ using a conjugate
gradient method. The resulting $T=0$ energies are 
compared in the following section.

The entropic sampling method was implemented generally as described in
section 2. The Metropolis algorithm \cite{mrrtt53} was used for
local updates.  We developed an initial estimate $J(E)$ for the
entropy $S(E)$ with a long run (approximately $10^7$ sweeps) starting
from a flat distribution: this initial estimate was used as the
starting distribution for the subsequent runs.  In three dimensions we
redid this initialization for each system size; in two dimensions we
initialized in this way for the 101 particle system and used
finite-size scaling\cite{sb95} from this distribution for
initialization at other system sizes.  Finite-size scaling does not
appear to work for glassy systems with quenched disorder\cite{spin1,spin3}.
We did not include this initial computer time in the comparisons: thus
we err on the side of entropic sampling.  Notice that in continuous
systems it is not obvious how to set the optimal bin size for the
histogram (unlike in spin systems, where the smallest energy step
determines the bin size).  We therefore tested several bin sizes.  For
the two--dimensional systems a fixed bin size of 0.001 has been used,
and in the three--dimensional systems, we found it useful to set the
bin size to $0.01/N$ energy units.  Our investigations suggest that
larger bin sizes can introduce artificial barriers in the low energy
range, and smaller bins lead to more noise.  To set a context, the
typical successful energy step in a 20 particle simulation in three
dimensions varied from around one at high temperatures to around 0.01
near the ground state.  We explored energy-dependent step sizes for
the single-atom moves, but they did not improve performance.  Entropic
sampling demands an upper cutoff for the energy: we use zero for the
upper limit for both two and three dimensions.

Simulated annealing uses locally the Metropolis update scheme with the
Boltzmann factor as the sample probability distribution.  The cooling
schedule implemented here is similar to the one used in references
\cite{wss87,ho94}, where the temperature is repeatedly lowered by a small
factor and then annealed.  In our runs, we choose fifty annealing
steps of 20,000 sweeps each, with an initial temperature of one and a
final temperature of 0.05; each temperature is thus cooled down by a
factor of $0.942$. The initial configurations were set at random.

The molecular dynamics routine used the velocity form of the Verlet
algorithm\cite{ta86}.  The unit of time is given by
$(m\sigma^2_{LL}/48\epsilon_{LL})^{1/2}$, where $m$ is the mass of the
particle: the Verlet time step in these units is $\delta t=0.01$.  The
system was coupled to a heat bath and the temperature was reduced
linearly in time according to $T_{\rm bath}~=~T_{\rm
start}~-~\gamma_{\rm MD} \times t$.  Note that the cooling here is
linear in time , as is traditional in molecular dynamics of
Lennard--Jones glasses\cite{v96}.  The cooling rate $\gamma_{\rm MD} =
1.0~10^{-4}$ was chosen so that the MD runs consume an amount of
computer time similar to that of the Monte Carlo algorithms.  This
cooling rate is in the middle of the range explored in recent
simulations, although our system sizes are much smaller\cite{vkb96}.
The initial configurations were equilibrated at a temperature $T_{\rm
start} = 1.0$ (at a small cost of computer time which we did not
factor into the comparisons), and cooled to the final temperature
$T_{\rm final}=0.05$, yielding approximately $2.0 \times 10^6$
molecular dynamics steps.

\section{Results}

In this section we will first compare the performance of the three
methods in locating low energy states of the two and
three--dimensional Lennard--Jones systems. The performance of the
three methods is remarkably similar. Second, we will compare low
energy configurations of the two--dimensional system to show that the
the algorithms get trapped in similar metastable states.  Third, we
will quantitatively analyze the trapping of the entropic sampling
algorithm in a metastable state.

We present the $T=0$ energies of the lowest energy configurations for
the three--dimensional Lennard--Jones systems in Table 1. For $N=20$
particles each algorithm is able to locate the same lowest energy
state, presumably the ground state. For $N=40$ and $N=60$ particles
the lowest energy state is found by entropic sampling. The gain in
energy $\Delta E_{40}$ over simulated annealing is around 0.03, and
the gain is 0.02 over molecular dynamics. For $N=80$ and $N=100$
particles the lowest energies are found by molecular dynamics, and the
gain over entropic sampling is $\Delta E_{80} \sim 0.05$ and $\Delta
E_{100}\sim 0.02$.

\vspace{0.3cm}

\noindent
{\footnotesize TABLE 1: $T=0$ energy per particle for the lowest energy
configuration found with entropic sampling, simulated annealing and
molecular dynamics.}

\setlength{\tabcolsep}{0.45cm}
\noindent
\begin{tabular}{|c||c|c|c|}  \hline
{\em N} & {\em Entropic}  &  {\em Simulated}  & {\em Molecular} \\ 
        & {\em Sampling}  &  {\em Annealing}  & {\em Dynamics} \\ \hline\hline
20      & -0.89 & -0.89  &  -0.89 \\ 
40      & -4.18 & -4.15  &  -4.16 \\ 
60      & -5.38  & -5.36  & -5.37 \\  
80      & -6.52  & -6.56  & -6.57 \\ 
100     & -6.85  & -6.86  &  -6.87 \\ \hline
\end{tabular}
\vspace{0.5cm}

There are three things to notice about this table. First, the dramatic
energy difference with increasing system size is due to the change in
the cutoff in the potential given by the minimum image rule. Using the
twenty particle cutoff in the larger system sizes, we found energies
which hardly varied with system size.  Second, the fact that these
energies differ in the third decimal place does not mean that the
differences are negligible. In ref. \cite{vkb96,v96} the dependence of
the final energy on the cooling rate for exactly this system was
studied using molecular dynamics: to gain an energy of $0.03$ starting
from the cooling rate we are using, they had to decrease the cooling
rate by a factor of ten. Third, as we argued in section three, any
gains given by entropic sampling disappear as the system size grows.

In Table 2 we list the mean and the standard deviation of the energies
from the thirty runs at each system size with each algorithm. It has
been found in the literature that the fluctuations for simulated
annealing are much larger than for entropic sampling \cite{ho94}.  We
find this to be true both for simulated annealing and for molecular
dynamics. Indeed, the average performance of entropic sampling remains
comparable to that of the other two methods, even at the larger system
sizes (where the extremal performance was worse).
\vspace{0.3cm}

\noindent
{\footnotesize TABLE 2: Mean energy per particle and the standard
deviation evaluated using all low energy configurations found by
entropic sampling, simulated annealing and molecular dynamics.}

\setlength{\tabcolsep}{0.3cm}
\noindent
\begin{tabular}{|c||c|c|c|}  \hline
{\em N} & {\em Entropic}  &  {\em Simulated}  & {\em Molecular} \\ 
        & {\em Sampling}  &  {\em Annealing}  & {\em Dynamics} \\ \hline\hline
20      & -0.84 $\pm$ 0.03  & -0.79 $\pm$ 0.06   & -0.81 $\pm$ 0.04  \\ 
40      & -4.10 $\pm$ 0.03  & -4.07 $\pm$ 0.05  & -4.08 $\pm$ 0.04   \\ 
60      & -5.34 $\pm$ 0.01  & -5.33 $\pm$ 0.03  & -5.33 $\pm$ 0.03   \\  
80      &  -6.50 $\pm$ 0.01  & -6.51 $\pm$ 0.02  & -6.51 $\pm$  0.03 \\ 
100     & -6.83 $\pm$ 0.01  & -6.84 $\pm$ 0.02  & -6.82  $\pm$ 0.02 \\ \hline
\end{tabular}
\vspace{0.5cm}

The Holy Grail of this field is to accelerate three--dimensional glass
simulations, bypassing barriers to relaxation. Perhaps this is too
high a standard --- nobody has such an algorithm. We now will apply
entropic sampling to a two--dimensional Lennard--Jones system, where
an effective cluster--flip acceleration method has been developed
\cite{wss87}.
In Tables 3 and 4 we show the extremal and the mean $T=0$ energies for
a variety of system sizes. Again, entropic sampling is slightly better
for the smaller systems, but the advantage disappears for the largest
system.

\vspace{0.3cm}
\noindent
{\footnotesize TABLE 3: Energy per particle for the lowest energy
configuration found with entropic sampling and simulated annealing.}

\setlength{\tabcolsep}{0.34cm}
\noindent
\begin{tabular}{|c||c|c|}  \hline
{\em N} & {\em Entropic Sampling}  & {\em Simulated Annealing} \\ \hline\hline
31      & -2.1  & -2.1  \\ 
66      & -2.27 & -2.22 \\ 
101     & -2.34 & -2.33 \\ 
160     & -2.37 & -2.38 \\ \hline
\end{tabular}
\vspace{0.3cm}

{\footnotesize TABLE 4: Mean energy per particle and the standard
deviation evaluated using all low energy configurations found by
entropic sampling and simulated annealing.}

\setlength{\tabcolsep}{0.33cm}
\noindent
\begin{tabular}{|c||c|c|}  \hline
{\em N} & {\em Entropic Sampling}  & {\em Simulated Annealing} \\ \hline\hline
31      & -1.97 $\pm$ 0.1  & -1.87 $\pm$ 0.17  \\ 
66      & -2.24 $\pm$ 0.04 & -2.12 $\pm$ 0.07  \\
101     & -2.29 $\pm$ 0.07 & -2.23 $\pm$ 0.06  \\
160     & -2.34 $\pm$ 0.05 & -2.34 $\pm$ 0.02  \\ \hline
\end{tabular}
\vspace{0.5cm}

It is remarkable how similarly the three different methods
perform. Although it seems to be known that molecular dynamics and
simulated annealing are comparable \cite{c97}, we are not aware of any
reference providing a direct comparison. Of course, comparisons of
efficiency are highly implementation dependent. The two Monte Carlo
methods could benefit from a temperature dependent step-size (although
we did experiment with it without finding any substantial
improvement).  One could refine the cooling schedule for the two
traditional methods.  One could introduce a temperature cutoff (like
in the original multicanonical approach) or use variable bin-sizes to
improve the entropic sampling method. Again, our experiments with
bin--size and cutoff were not encouraging. Our main conclusion is that
the choice of methods is a matter of taste. In particular we are
encouraged by the fact that Monte Carlo methods are competitive,
especially as the adapt easily to cluster acceleration methods.

All three methods suffer from the large number of metastable states
prevalent in the configuration space of the two and three--dimensional
systems, and thus are not capable of locating ground states. To show
that they find similar metastable states, we plot in figures 3 and 4
the lowest energy configuration found by entropic sampling and by
simulated annealing. 
\begin{figure}[htbp]
\begin{center}
\leavevmode
\epsfysize=6cm \epsfbox{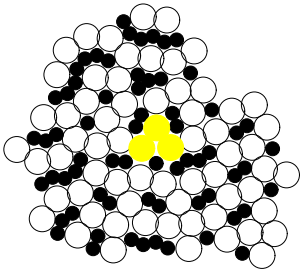}
\caption{ Lowest energy configuration generated with entropic sampling. The grey particles forming a triangle represent a defect.}
\end{center}
\label{fig3}
\end{figure}
%FIGURE 3+4
\begin{figure}[htbp]
\begin{center}
\leavevmode
\epsfysize=6cm \epsfbox{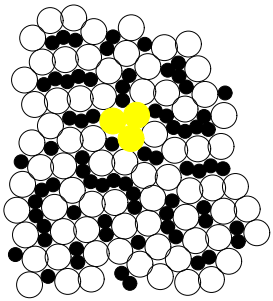}
\caption{ Lowest energy configuration generated with simulated annealing.The grey particles forming a triangle represent a defect.}
\end{center}
\label{fig4}
\end{figure}

These configurations are typical representatives of metastable states
for the two--dimensional system. The defects are clusters of three
large particles, which are shown in grey in figures 3 and 4.

Why is entropic sampling not bypassing the free--energy barriers to
relaxation? We finish this section with a vivid illustration of how
the entropic sampling algorithm gets trapped in a metastable state.
In the bottom half of figure~5 we plot the energy as a function of
time: at very short times it performs a random walk in energy space as
advertised, but it rapidly gets trapped in a low energy metastable
state. 

The simulation shown in figure~5 is the same as the runs for $N=100$
particles tabulated in Tables 1 and 2 except for two important
differences. (1)~The runs of Table 1 and 2 ran for $10^6$ sweeps, here
we ran for $10^7$ sweeps. (2)~The entropy estimate $J(E)$ for the runs
in Tables 1 and 2 was dynamically updated every $10^5$ sweeps using
the recursive updating scheme equation~(2). Here we calculated a best
estimate $\overline{J(E)}$ from the thirty runs in the Tables and used
this function as a fixed entropy estimate.  The best estimate
$\overline{J(E)}$ (comprising information from $40 \times 10^6$
sweeps) is a sufficiently smooth function that we don't expect (or
observe) the system to be trapped in some artificial well resulting
from statistical fluctuations in $J(E)$.

%FIGURE 5
\begin{figure}[htbp]
\begin{center} \leavevmode
\epsfysize=7cm \epsfbox{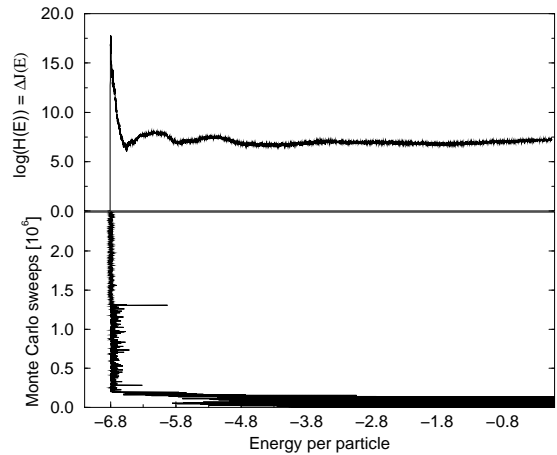}
\caption{
An entropic sampling run, showing trapping into a metastable state.
In this run, we used a fixed $\overline{J(E)}$ ({\it i.e.} no
dynamical updating) gleaned from several previous runs of the same
system.  {\bf Energy vs.~time} is shown on the bottom panel. Only the
first quarter of the time series is shown.  At short times, we observe
a random walk in energy.  At times below $2 \times 10^5$ sweeps the system
randomly walks through energies above the glass transition. The system
then gets trapped in two successively lower metastable free--energy
wells (see figure~6). {\bf The Histogram} $\log(H(E)) = \Delta J(E)$
of visited states measured during the simulation is shown in the upper
panel. Note the large peak associated with the trapping. The second
and third peak are above the glass transition and correspond to more
transient states.}
\end{center}
\label{fig5}
\end{figure}

The top half of figure~5 shows the logarithm of the histogram $H(E)$
tabulating the visited states as a function of energy. This function
is important as it is used in the recursive updating scheme
$\log(H(E)) = \Delta J(E) = J(E)_{\rm update} - J(E)_{\rm estimate}$
(see equation~(2)).

Figure~6 shows that the system is trapped in a single harmonic
metastable state. The upper panel shows an expanded view of the first
peak in $\Delta J(E)$. For times after $1.3 \times 10^6$ the system
exclusively samples in a single well: repeated quenches yield the same
minimum energy $E_1 = -6.8241$. This is a metastable state: as seen in
Table 1 the true ground state has an energy $\leq -6.87$. 

%FIGURE 6
\begin{figure}[htbp]
\begin{center} \leavevmode
\epsfysize=7.5cm \epsfbox{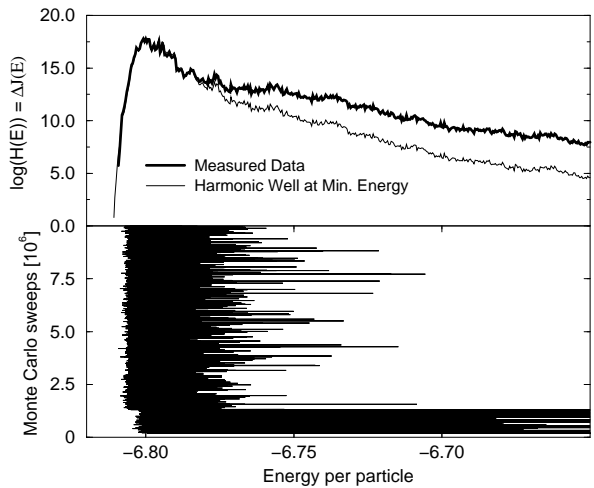}
\caption{
Expanded view of figure~5 at low energies.  {\bf Energy vs. time} is
 shown on the bottom panel. From $2 \times 10^5$ to $1.3 \times 10^6$
 sweeps, the system seems trapped in a number of states: repeated
 quenches yield metastable energies clustering around $E_2 = -6.82\pm
 0.002$.  The system then falls into a slightly lower state, with
 energy $E_1 = -6.8241$; repeated quenches show that it stays in this
 single well for the remainder of the simulation.  The equilibration
 within this lowest well is excellent: the acceptance ratio is near
 50\% and the system exhibits an efficient random walk in energy
 within the single well.  {\bf The Histogram} $\log(H(E)) = \Delta
 J(E)$ of visited states is shown in the upper panel as a dark
 line. The light line is the theoretical prediction assuming a single,
 harmonic well at energy $E_1 = -6.8241$: $\Delta J_{\rm harmonic}(E)
 = (3 N/2 - 1) \log(E-E_1) - \overline{J(E)} + C$.  The bumps in the
 theoretical curve are due to the irregularities in our initial
 estimate $\overline{J(E)}$; the identical--looking bumps in the
 measured data reflect the effects of $\overline{J(E)}$ in weighting
 the histogram.  Our theoretical prediction describes the measured
 data well for the energy range explored during the last portions of
 the simulation.  The range above $E/N\sim -6.78$ is underestimated as
 in this region the data is composed largely from states corresponding
 to the minima clustered around $E_2$.}
\end{center}
\label{fig6}
\end{figure}

In the harmonic approximation we can analytically calculate the
density of states and compare the contribution from the metastable
state directly to the measured data.  The harmonic density of states
has the form
\begin{equation}
\Omega_{\rm harmonic}(E) \propto \left({2(E-E_1) \over K}\right)^{{3N \over 2} - 1} \; ,
\end{equation}
where $K$ involves the geometric mean of the phonon frequencies and
can be thought of as a typical spring constant. In the entropic
sampling algorithm the probability of sampling a state in this
harmonic well is given by the ratio of the density of states in the
single well divided by the estimated density of states
$\exp(\overline{J(E)})$. This probability is compared directly to the
histogram of sampled states in the upper half of figure~6.  The system
is trapped in a single harmonic well. 

By running for shorter times and by dynamically updating the entropy
estimate during each run, we have substantially mitigated the trapping
problem of entropic sampling shown in figures 5 and 6. Imagine the
entropy estimate $J(E)$ after updating it by adding $\log(H(E))$. The
acceptance ratio to leave the region given by $1/J(E)$ will
dramatically increase and thus the trapping will be bypassed, as
indeed we observed in practice. Lingering near a state increases the
estimate entropy in that region and eventually push the system
out. But dynamical updating should not be an essential ingredient of
the algorithm, which is formulated presuming an {\it a priori}
knowledge of the entropy as a function of energy. Formally dynamical
updating violates the Markovian character of the algorithm and
convergence to the equilibrium state is no longer guaranteed. In
practical terms it is very distressing that the algorithm needs to
produce a noticeable bump in the density of states to escape from a
metastable state. The comparison against molecular dynamics and
simulated annealing would be substantially more unfavorable for long
runs without dynamical updating.

The figures 5 and 6 are a tangible illustration of the trapping
mechanism depicted in figures 1 and 2.  The inaccessible metastable
states contributing to $\overline{J(E)}$ form a strange type of
entropic barrier around the metastable state $E_1$. Leaving $E_1$ via
a saddlepoint at $E_2 > E_{\rm peak} \sim -6.8$ is suppressed by roughly
the exponential of $\Delta J(E_2) - \Delta J(E_{\rm peak})$ shown in
figure~6.

Slow cooling in molecular dynamics or simulated annealing can lead to
trapping in metastable states due to large energy barriers. Entropic
sampling and the other multicanonical methods get trapped in
metastable states because of large entropic barriers imposed by the
algorithm. In both case the algorithms are sabotaged by the large
number of low lying metastable states. Entropic sampling provides a
new insight into this problem but doesn't provide a solution.

\section{Conclusions}

In this study we applied the multicanonical method entropic sampling
to Lennard--Jones systems. We focused on the ability of the algorithm
to find ground states of these glassy systems and compared the
performance to the two traditional glassy simulation methods simulated
annealing and molecular dynamics. The use of entropic sampling didn't
reveal any new insights into the ground state properties of
Lennard--Jones glasses. We explain these results on the basis of the
following observations.

First, in the thermodynamic limit multicanonical methods are locally
equivalent to simulated annealing. Furthermore the global dynamics of
multicanonical sampling resembles a random heating and cooling of the
sample. Thus for large systems simulated annealing and multicanonical
sampling must have the same properties. In principle multicanonical
sampling has the advantage of providing the density of states, which
allows to evaluate the canonical distribution function. In glasses
this feature is not necessarily helpful, as the multicanonical methods
samples phase space as slowly as the annealing methods, thus in
practice multicanonical sampling will not be able to extract any
equilibrium expectation values better than simulated annealing.

Second, the large number of inaccessible metastable states imposes a
bizarre entropy barrier to the multicanonical method. The algorithm
simply gets stuck in a metastable state, as it might using molecular
dynamics and simulated annealing. We underlined this point by
comparing the probability distribution estimated by the algorithm
inside the metastable state with a theoretical expression derived in
the harmonic approximation.

Furthermore our results emphasize the known fact, that simulated
annealing and molecular dynamics have similar performance in glassy
systems. As a consequence one should acknowledge the importance of
averaging over many molecular dynamics trajectories especially for
glassy systems. Averages over an ensemble of trajectories are a basic
concept in Monte Carlo simulations, the striking similarity in
performance to molecular dynamics simulations is a hint to the
importance of similar averages in glassy molecular dynamics
simulations.

Finally, the goal of finding a method which gains an exponential
speed--up of glassy simulations still remains. Our study clearly
indicates that standard reweighting techniques will presumably be no
substantial help in tackling this problem. The complicated structure
of the glassy configuration space needs more intelligent algorithms,
which are not only able to bypass energy barriers but also to find an
efficient path through the rugged energy landscape.

\begin{acknowledgments}
We would like to thank M. E. J. Newman, J. Jacobsen, K. W. Jacobsen,
G. Chester and R. Kree for many useful and illuminating discussions,
and B. Berg for very helpful and elucidating comments on various
points. The work of JPS was supported by NSF Grant DMR-9419506. KKB is
grateful for support by the German Academic Exchange Service
(Doktorandenstipendium HSP II/AUFE). Computational support was
provided by the Cornell Theory Center.
\end{acknowledgments}

\end{document}